\RequirePackage{ifpdf}
\ifpdf 
\documentclass[pdftex]{sigma}
\else
\documentclass{sigma}
\fi

\newcommand{\norm}[1]{\left\Vert {#1} \right\Vert}
\newcommand{\abs}[1]{\left\vert {#1} \right\vert}
\newcommand{\set}[1]{\left\{ {#1} \right\}}
\newcommand{\parenthese}[1]{\left( {#1} \right)}
\newcommand{\dv}[2]{\frac{\mathrm d{#1}}{\mathrm d{#2}}}

\newcommand{\setN}{{\mathbb N}}

\newcommand{\setR}{{\mathbb R}}

\newcommand{\Lequi}{\Longleftrightarrow}
\newcommand{\precc}{\prec\!\!\prec}

\begin{document}

\allowdisplaybreaks

\renewcommand{\thefootnote}{$\star$}

\renewcommand{\PaperNumber}{064}

\FirstPageHeading

\ShortArticleName{Global Eikonal Condition for Lorentzian Distance Function}

\ArticleName{Global Eikonal Condition for Lorentzian Distance \\ Function in Noncommutative Geometry\footnote{This paper is a
contribution to the Special Issue ``Noncommutative Spaces and Fields''. The
full collection is available at \href{http://www.emis.de/journals/SIGMA/noncommutative.html}{http://www.emis.de/journals/SIGMA/noncommutative.html}}}

\Author{Nicolas FRANCO}

\AuthorNameForHeading{N.~Franco}

\Address{GAMASCO, Department of Mathematics, University of Namur FUNDP,\\ 8 Rempart de la Vierge, B-5000 Namur, Belgium}
\Email{\href{mailto:nicolas.franco@math.fundp.ac.be}{nicolas.franco@math.fundp.ac.be}}

\ArticleDates{Received March 30, 2010, in f\/inal form August 06, 2010;  Published online August 17, 2010}

\Abstract{Connes' noncommutative Riemannian distance formula is constructed in two steps, the f\/irst one being the construction of a path-independent geometrical functional using a global constraint on continuous functions. This paper generalizes this f\/irst step to Lorentzian geometry. We show that, in a globally hyperbolic spacetime, a single global timelike eikonal condition is suf\/f\/icient to construct a path-independent Lorentzian distance function.}


\Classification{58B34; 53C50}

\section{Introduction}

\looseness=1
In noncommutative geometry \cite{C94,MC08}, it is well known that a spectral triple carries a notion of distance. In particular, if this spectral triple is associated to a Riemannian spin manifold, then the distance function coincides with the usual geodesic distance. However, this correspondence is limited to the Riemannian side of geometry, and there is at this time no complete counterpart to deal with pseudo-Riemannian geometry (although there are several proposals, e.g.~\cite{Mor,Pas,Stro}), and in particular Lorentzian geometry, which is just the geometry of our physical spacetime.

\looseness=1
The relation between spectral triples and Riemannian geometry arises from a distance function established by Connes in the nineties \cite{C94,C96}. This distance function is constructed in two steps. At f\/irst a geometrical functional is set, giving the distance between two points on a~mani\-fold $M$ without reference to any path, by use of a global constraint on $C(M)$. Then this constraint is translated into an algebraic formalism and so can be extended to a noncommutative framework. We will see that the f\/irst step of the construction of such distance function has a~corresponding one in Lorentzian geometry, if we work under the condition of global hyper\-boli\-city.

\looseness=1
So this work is a f\/irst step in the construction of a noncommutative Lorentzian distance. The result will be similar on some points to the function constructed by Moretti in \cite{Mor} but with signif\/icant dif\/ferences. Indeed, if Moretti proposes a constraint based on certain causal sets, we will show that a single global eikonal inequality based on the gradient is suf\/f\/icient to contain all informations. The way to obtain the distance function diverges because we will try to follow the same construction as the Riemannian one. Moreover, the global character of the condition makes this distance function a suitable candidate for a generalization in noncommutative spaces.

\section{Review of the construction of the Riemannian distance\\ in noncommutative geometry}

At f\/irst, we will focus on the Riemannian distance formula developed by Alain Connes \cite{C94,C96} and review the common way to establish it~\cite{Var}.

Let us consider a compact Riemannian manifold $(M,g)$ (supposed connected) and two points $p$ and $q$ on it. Choose an arbitrary piecewise smooth curve $\gamma : [0,1] \rightarrow M$ with $\gamma(0) = p$ and $\gamma(1) = q$.

Then, for each function $f \in C^\infty(M)$,
\begin{gather*}
f(q) - f(p) = f(\gamma(1)) - f(\gamma(0)) = \int_0^1 \dv{}{t} f(\gamma(t)) \, dt = \int_0^1 df(\dot\gamma(t)) \, dt = \int_0^1  g( \nabla f,\dot\gamma(t)) \, dt.
\end{gather*}
By use of the Cauchy--Schwartz inequality:
\begin{gather*}
\abs{ f(q) - f(p) } \leq \int_0^1  \abs{g( \nabla f,\dot\gamma(t))} \, dt \leq \int_0^1  \abs{ \nabla f} \abs{\dot\gamma(t)} \, dt \\
\phantom{\abs{ f(q) - f(p) }}{}
\leq \norm{  \nabla f}_\infty  \int_0^1  \abs{\dot\gamma(t)} \, dt = \norm{  \nabla f}_\infty l(\gamma),
\end{gather*}
where $l(\gamma)$ denotes the length of the curve. So the condition $\norm{  \nabla f}_\infty \leq 1$ is enough to get $\abs{ f(q) - f(p) } \leq d(p,q) = \inf\set{ l(\gamma) : \gamma \text{ piecewise smooth curve from $p$ to $q$}}$. In fact, the condition $\norm{  \nabla f}_\infty \leq 1$ can be replaced by a larger condition $\text{ess} \sup \norm{  \nabla f} \leq 1$ (essential supremum)  which allows us to work with the set $\mathcal A \subset C(M)$ of Lipschitz functions on $M$. Those functions are almost everywhere dif\/ferentiable  (a.e.~dif\/ferentiable), that means that we work with continuous functions such that $ \nabla f$ is bounded and def\/ined everywhere, except on a set of measure zero.

 Then we can set
\[
d(p,q) = \sup\set{  \abs{f(q)-f(p)} \ :\  f \in \mathcal A,\  \text{ess} \sup \norm{  \nabla f} \leq 1 }
\]
which would correspond to the usual Lorentzian distance if we can f\/ind a function $f\in\mathcal A$ giving $\abs{f(q)-f(p)} = d(p,q)$. The equality function is just the usual distance as function of its second argument $f(\cdot) = d_p(\cdot) = d(p,\cdot)$. Indeed, $d_p$ is Lipschitz, with $\norm{  \nabla d_p} = 1$ except on a set of measure zero (called the cut locus).

The particularity of this formula is that there is no reference to any path on the manifold $M$, despite the fact that the def\/inition of the usual distance depends explicitly on paths.

The next step is the translation of this formula into a noncommutative formalism. If $M$ is a~spin manifold, we have
\[
\text{ess} \sup \norm{  \nabla f} \leq 1\ \Lequi\ \norm{[D,f]} \leq 1
\]
 with $D$ the Dirac operator  \cite{C94,Var}. The f\/inal distance formula is
\[
d(p,q) = \sup\set{   \abs{f(q)-f(p)} \ :\  f \in \mathcal A,\  \norm{[D,f]}  \leq 1 }
\]
and can be extended to noncommutative spaces, by replacing points on a geometrical manifold~$M$ by states on a noncommutative algebra.

\section{Lorentzian distance and the timelike eikonal inequality}

From this point, we will switch from Riemannian manifolds to Lorentzian ones. We will assume that the reader is familiar with Lorentzian geometry, so we will only give def\/initions and pro\-per\-ties which are non usual or of particular interest. For the remaining, the reader can refer to~\cite{Beem,ONeill}. We will use $p \preceq q$ to indicate that $p = q$ or there is a future directed causal curve from~$p$ to~$q$, and $p \precc q$ to indicate that there is a future directed timelike curve from~$p$ to~$q$. $p \prec q$ means that $p \preceq q$ and $p\neq q$. The signature of the metric will be $(-,+,+,+)$.

\begin{definition}
The Lorentzian distance function on an oriented Lorentzian manifold $(M,g)$ is the function $d : M \times M \rightarrow [0,+\infty) \cup \set{+\infty}$ def\/ined by
\begin{gather*}
d(p,q) =
\begin{cases}
 \sup\{ l(\gamma) : \gamma \text{ future directed causal piecewise} \ & \ \\
  \qquad \text{smooth curve from $p$ to $q$}\} &\text{if } p \prec q,\\
 0 &\text{if } p \nprec q,
\end{cases}
 \end{gather*}
 with $l(\gamma) = \int \sqrt{ -g( \dot\gamma(t) ,  \dot\gamma(t) )}\ dt$ the length of the curve.
\end{definition}

Lorentzian geometry is really dif\/ferent from Riemannian one on lots of aspects. For example, the distance between to points is no more commutative, and is in fact a function which is not null only on a single future cone. Any formula for Lorentzian distance must respect these aspects. Practically, taking two arbitrary distinct points $p$ and $q$, we have to dissociate those three cases:
\begin{itemize}\itemsep=0pt
\item $p \prec q$, where $p$ and $q$ are causally related, with $q$ in the future of $p$;
\item $q \succ p$, where $p$ and $q$ are causally related, with $q$ in the past of $p$;
\item $p \nprec q$ and $q \nprec p$, where $p$ and $q$ are not causally related.
\end{itemize}

Moreover, usual inequalities as triangle or Schwartz do not hold in Lorentzian geometry, but we have inverse counterparts:

\begin{lemma}
In Lorentzian geometry, we have the following inequalities:
\begin{enumerate}\itemsep=0pt
\item if $p \preceq q \preceq r$, the inverse triangle inequality holds:
\[
d(p,q) + d(q,r) \leq d(p,r),
\]
\item if $v$ and $w$ are timelike vectors, the inverse Cauchy--Schwartz inequality holds:
\[
\abs{g(v,w)} \geq \sqrt{- g(v,v)} \sqrt{- g(w,w)}.
\]
\end{enumerate}
\end{lemma}

\begin{proof}
Proofs can be found in \cite{Beem,ONeill}.
\end{proof}

Now we will try to adapt the Riemannian distance function by following the same procedure than above.

Let us consider a 4-dimensional oriented Lorentzian manifold $(M,g)$ and two points~$p$ and~$q$ on it such that $p \precc q$. Choose an arbitrary future directed timelike piecewise smooth curve $\gamma : [0,1] \rightarrow M$ with $\gamma(0) = p$ and $\gamma(1) = q$ (there must exist at least one such curve because $p \precc q$). Then, for each function $f \in C^\infty(M,\setR)$:
\begin{gather*}
f(q) - f(p) = f(\gamma(1)) - f(\gamma(0)) = \int_0^1 \dv{}{t} f(\gamma(t)) \, dt = \int_0^1 df(\dot\gamma(t)) \, dt = \int_0^1  g( \nabla f,\dot\gamma(t)) \, dt.
\end{gather*}
Notice that $\dot\gamma(t)$ is everywhere a future directed timelike vector. If we suppose that $\nabla f$ is everywhere timelike with constant orientation (we will take past directed), then $g( \nabla f,\dot\gamma(t))$ is of constant sign (hence positive):
\[
f(q) - f(p) = \int_0^1  g( \nabla f,\dot\gamma(t)) \, dt = \int_0^1 \abs{ g( \nabla f,\dot\gamma(t)) }\, dt.
\]

We can apply the inverse Cauchy--Schwartz inequality:
\begin{gather*}
\int_0^1  \abs{g( \nabla f,\dot\gamma(t))} \, dt  \geq \int_0^1  \sqrt{-g(\nabla f,\nabla f)} \sqrt{-g(\dot\gamma(t),\dot\gamma(t))} \, dt  \geq  \inf\set{{\sqrt{-g(\nabla f,\nabla f)}}}\; l(\gamma).
\end{gather*}

Now we can see that this construction holds also if $\gamma$ is causal instead of just timelike. Indeed, $g( \nabla f,\dot\gamma(t))$ is non-negative for a future directed lightlike vector $\dot\gamma(t)$ and Cauchy--Schwartz inequality becomes $g( \nabla f,\dot\gamma(t)) \geq 0$. So we can extend our construction to the case $p \prec q$.

Now we can see that $f(q) - f(p) \geq d(p,q)$ holds if we impose the following conditions on $f$:
\begin{itemize}\itemsep=0pt
\item  $\sup\, g( \nabla f, \nabla f ) \leq -1$;
\item $\nabla f$ is past directed.
\end{itemize}

The f\/irst condition is a global eikonal timelike condition. Once more, this condition is totally independent of any path on the manifold, so could give rise to a translation into an algebraic framework in order to be extended to noncommutative spaces. Of course this generalization is far from trivial and does not enter the scope of this paper.

The second condition is in reality not a restrictive condition but a simple choice of time orientation, which can be easily translated into a noncommutative framework. Indeed, functions obeying the f\/irst condition can be easily separated in two sets. To do that, just take two f\/ixed points $p_0$ and $q_0$ (or the corresponding states in a noncommutative algebra) such that $\inf_f \abs{ f(q_0) -f(p_0) } > 0$. Our two sets are
\begin{gather*}
\set{ f : g( \nabla f, \nabla f ) \leq -1,\; f(q_0) - f(p_0) > 0}\!\!\quad\text{and}\quad\!\!  \set{ f : g( \nabla f, \nabla f ) \leq -1,\; f(q_0) - f(p_0) < 0}\!
\end{gather*}
 and the choice of one of them corresponds to the choice of a particular time orientation.

Once more, we can extend the set of functions $f \in C^\infty(M,\setR)$ to a larger set, including continuous functions a.e.\ dif\/ferentiable, and use the weaker condition $\text{ess}\sup\, g( \nabla f, \nabla f ) \leq -1$. Nevertheless, due to the Lorentzian characteristic of the space, Lipschitz functions are not suitable. We must extend to an even larger set which we will denote here by $\mathcal A$. In fact, $\mathcal A$ must be a set of continuous functions on $M$ which are a.e.\ dif\/ferentiable and respect the fundamental theorem of calculus at least on causal paths (absolutely continuous functions).

There are many ways to characterize such set $\mathcal A$, and the choice of a particular one should take in account the generalization to noncommutative spaces. One can use the $\lambda$-absolutely continuous condition developed mainly by J.~Maly and S.~Hencl for functions of several variables or use Sobolov spaces $W^{1,p}(M)$ ($p>4$)~\cite{Hen,Mal}.

Another possibility for the space $\mathcal A$ is the space of continuous causal functions, i.e.\ functions which do not decrease along every causal future-directed curve. Indeed, by Lebesgue dif\/ferentiation theorem, such functions are of bounded variation on any future directed timelike smooth curve $\gamma : [0,1] \rightarrow M$, so a.e.\ dif\/ferentiable, and the fundamental theorem turns to be an inequality:
\[
f(q) - f(p) = f(\gamma(1)) - f(\gamma(0)) \geq \int_0^1 \dv{}{t} f(\gamma(t)) \, dt
\]
and the above construction remains valid.

From now, and for the rest of this paper, we will f\/ix $\mathcal A \subset C(M,\setR)$ as the space of continuous causal functions on $M$. Then we present our principal theorem, which takes place in a Lorentzian manifold with a globally hyperbolic condition:

\begin{theorem}\label{mainth}
Let $(M,g)$ be a globally hyperbolic spacetime and $d$ the Lorentzian distance function on $M$, then
\[
d(p,q) = \inf\set{ \langle f(q)-f(p) \rangle \ :\  f \in \mathcal A,\  \text{\rm ess} \sup g( \nabla f, \nabla f ) \leq -1, \ \nabla f \text{ past directed}}
\]
where $\langle \alpha \rangle = \max\set{0,\alpha}$.

\end{theorem}

In the next section, we give a proof of this theorem with $\mathcal A$ being the space of continuous causal functions, but the result is still valid with any suitable space $\mathcal A$, provided that the Lorentzian distance function belongs to $\mathcal A$ as a function of its second argument. The globally hyperbolic condition is a useful tool in order to construct functions giving the equality. The reason is that a global eikonal condition can be obtained only if we have a global knowledge on the causal structure of the manifold, and the globally hyperbolic condition does so.

\section{Proof of the Lorentzian distance function}

In the following, we will work in a globally hyperbolic spacetime $(M,g)$ (necessarily oriented). Our goal is to prove mathematically the Theorem \ref{mainth} above. We have already obtained the result:

\begin{lemma}\label{lemmainequa}
If $p \prec q$, then
\begin{gather*}
\forall \, f \in \mathcal A\text{ such that } \text{\rm ess} \sup g( \nabla f, \nabla f ) \leq -1   \text{ and }  \nabla f \text{ is past directed}, \  f(q)-f(p)   \geq  d(p,q).
\end{gather*}
\end{lemma}

We will separate the remaining in three cases:
\begin{itemize}\itemsep=0pt
\item if $p \prec q$, we will show that there exists a sequence of functions $f_\epsilon \in\mathcal A$ ($\epsilon > 0$) respecting the two given conditions and such that $d(p,q) \leq f_\epsilon(q) -f_\epsilon(p) < d(p,q) + \epsilon$ (Lemma \ref{equality});
\item if $p \succ q$, then there exists a function $f \in\mathcal A$ respecting the two given conditions and such that \mbox{$f(q)-f(p) \leq 0$} (Corollary \ref{equalityrev});
\item if $p \nprec q$ and $q \nprec p$, then there exists a sequence of functions $f_\epsilon \in\mathcal A$ ($\epsilon > 0$) respecting the two given conditions and such that $\abs{ f_\epsilon(q) -f_\epsilon(p) } < \epsilon$ (Lemma \ref{zero}).
\end{itemize}

With these three results, the proof of the Theorem \ref{mainth} will be completed, but we need to start f\/irst with some technical lemmas.

\begin{lemma}
In Lorentzian geometry, we have the following properties:
\begin{enumerate}\itemsep=0pt
\item if $(M,g)$ is globally hyperbolic, then $d$ is a finite continuous function on $M \times M$;

\item $d_p = d(p,\cdot) $ satisfies the timelike eikonal equation $g( \nabla d_p , \nabla d_p) = -1$ on $I^+(p) \setminus C^+(p)$ where $C^+(p)$ is called the cut locus of $p$;

\item if $(M,g)$ is globally hyperbolic, the cut locus  $C^+(p)$ has measure zero;

\item  if $(M,g)$ is globally hyperbolic, then $d_p\in\mathcal A$.
\end{enumerate}
\end{lemma}

\begin{proof}
Proof of (1) can be found in \cite{Beem,ONeill}, proof of (2) in~\cite{EGK} and (3) can be found in~\cite{Mor}. By def\/inition of the distance, $d_p$ is trivially non-decreasing along every causal future-directed curve.
\end{proof}

The basic idea of our construction is to create some functions with a dif\/ferent behaviour in two regions:
\begin{itemize}\itemsep=0pt
\item the f\/irst region is a region containing the points $p$ and $q$ where these functions correspond to a simple suitable distance function;
\item the second region is the remaining of the manifold where these functions are locally f\/inite sums of distance functions in order to have the eikonal condition respected in the whole space.
\end{itemize}

The next lemma will be useful to construct the second region.

\begin{lemma}\label{lemmaprincipal}
Let $S$ be a smooth spacelike Cauchy surface and two points $q \in J^+(S)$ $($resp.\ $q \in J^-(S))$ and $q' \in I^+(q)$ $($resp.\ $q' \in I^-(q))$. There exists a~function $f \in \mathcal A$ $($resp.\ $-f \in \mathcal A)$ such that:
\begin{itemize}\itemsep=0pt
\item $f \geq 0 $;
\item $g( \nabla f, \nabla f ) \leq -1$ and $\nabla f$ is past directed (resp.\ future directed) where $f > 0$, except on a~set of measure zero $($so $f$ is a.e.\ differentiable$)$;
\item $f > 0$ on $J^+(S) \setminus I^-(q')$ $($resp.\ on $J^-(S) \setminus I^+(q'))$;
\item $f = 0$ on $J^-(q)$ $($resp.\ on $J^+(q))$.
\end{itemize}
\end{lemma}

\begin{proof}
At f\/irst, notice that $J^-(q) \subset  I^-(q')$. Indeed, $q \precc q'$, so $I^-(q) \subset  I^-(q')$. To verify that $J^-(q) = \overline{I^-(q) }\subset  I^-(q')$ consider that $I^-(q') = \set{ p\in M : d(p,q') > 0}$ as shown in \cite{Beem} and that for every point $ z\in J^-(q)$, $d(z,q') \geq d(z,q) + d(q,q') > 0$.

On the smooth spacelike surface $S$ we can consider the Riemannian metric $g_R$ which is the restriction of the global metric $g$ to $S$, with a Riemannian distance $d_R$ and a topology associated. If we take a point $p \in I^-(S)$, the intersection $I^+(p) \cap S$ is an open subset of $S$ (in the topology of $S$) with a f\/inite diameter $d_R(I^+(p) \cap S) < \infty $  because $J^+(p) \cap J^-(S)$ is compact (see \cite{Wald}). We will work with points $p$ closed to the surface $S$ such that $d_R(I^+(p) \cap S)$ is small. Let us def\/ine
\[
P = \set{ p \in I^-(S) \setminus J^-(q) : d_R(I^+(p) \cap S) < 1}.
\]

Then the collection
\[
W = \set{ I^+(p) \cap S : p \in P}
\]
is an open covering of the closed set $S \setminus I^-(q')$ (because we have $I^-(q') \supset J^-(q)$). We will show that there exists a locally f\/inite subcovering by a method similar to \cite{BS03}. Let us f\/ix $s \in S$ and consider open and closed balls $B_s(r)$ and $\bar B_s(r)$ in~$S$ of center $s$ and radius $r$ for the distance~$d_R$. The following subsets are compact in $S$:
\[
S_n = \bar B_s(n) \setminus \parenthese{ B_s(n-1) \cup I^-(q')}, \quad n \in \setN \qquad \text{with} \qquad \bigcup_{n\in \setN} S_n = S \setminus I^-(q').
\]

For each $S_n$ we can f\/ind a f\/inite subset $\set{W_{1n}, \dots, W_{k_n n}} \subset W$ which covers $S_n$. Then $W' = \set{W_{kn} : n\in \setN , k = 1, \dots, k_n}$ is a locally f\/inite subcovering of $S \setminus I^-(q')$ because every $W_{kn}$ has diameter smaller that 1. So there exists a subset of points $P' \subset P$ such that $W' = \set{ I^+(p) \cap S : p \in P'}$ is locally f\/inite.

From that, we can show that $\set{I^+(p) : p \in P'}$ is a locally f\/inite covering of  $J^+(S) \setminus I^-(q')$. Indeed, take a point $z$ in $J^+(S) \setminus I^-(q')$. The set $I^-(z) \cap S$ is an open set of f\/inite diameter in $S$ which intersects only a f\/inite number of $W_{kn}$ (and must intersect at least one because $S \setminus I^-(q')$ is a Cauchy surface for $M \setminus I^-(q')$) and so $I^-(z)$ contains a non empty but f\/inite subset of $P'$. The same reasoning can be done for a small neighbourhood of $z$.

Now we can construct the non-negative function:
\[
f(z) = \sum_{p \in P'} d(p,z) = \sum_{p \in P'} d_p(z).
\]
This function is well def\/ined because the sum is pointwise f\/inite and is continuous by continuity of the distance function. $f$ is null on $J^-(q)$ because no $p \in P'$ belongs to $J^-(q)$ and is positive on $J^+(S) \setminus I^-(q')$ because every point in $J^+(S) \setminus I^-(q')$ is strictly inside the light cone of at least one $p \in P'$. For every $z \in M$ we can f\/ind a neighbourhood where $f$ is  a f\/inite sum $\sum d_p$ of distance functions, so $f\in\mathcal A$. Because $\nabla d_p$ is well def\/ined except on a set of measure zero, by countability of the measure, the locally f\/inite sum $\nabla f = \sum \nabla d_p$ is well def\/ined except on a~set of measure zero. Then where $\nabla f$ is well def\/ined and $f$ positive, we have that $\nabla f$ is timelike past directed (because it is the sum of null or timelike past directed vectors), and
\[
g( \nabla f, \nabla f ) =  \sum_{p} g( \nabla d_p, \nabla d_p ) +  2 \sum_{p \neq p'} g( \nabla d_p, \nabla d_{p'} ) \leq -1,
\]
where the f\/irst sum contains terms equal to $-1$ or $0$, with at least one term equal to $-1$, and the second sum contains terms negative or null because all $\nabla d_p$ are null or timelike past directed.

In the reverse case, ($q \in J^-(S)$ and $q' \in I^-(q)$), we can do an identical proof by reversing future and past sets and by taking $f(z) = \sum_{p \in P'} d(z,p)$ as a function with null or timelike future oriented gradient which is non-increasing along every causal future-directed curve.
\end{proof}

\begin{corollary}\label{corollaryprincipal}
Let $S$ be a smooth spacelike Cauchy surface and four points $q_1,q_2 \in J^+(S)$ and $q_1' \in I^+(q_1)$, $q_2' \in I^+(q_2)$ $($resp.\ $q_1,q_2 \in J^-(S)$, $q_1' \in I^-(q_1)$, $q_2' \in I^-(q_2))$. There exists a~function $f \in \mathcal A$ $($resp. $-f \in \mathcal A)$ such that:
\begin{itemize}\itemsep=0pt
\item $f \geq 0 $;
\item $g( \nabla f, \nabla f ) \leq -1$ and $\nabla f$ is past directed $($resp.\ future directed$)$ where $f > 0$, except on a~set of measure zero;
\item $f > 0$ on $J^+(S) \setminus \parenthese{I^-(q_1') \cup I^-(q_2')}$ $($resp.\ on $J^-(S) \setminus \parenthese{I^+(q_1') \cup I^+(q_2')})$;
\item $f = 0$ on $J^-(q_1) \cup J^-(q_2)$ $($resp.\ on $J^+(q_1) \cup J^+(q_2))$.
\end{itemize}
\end{corollary}

\begin{proof}
The proof is identical to Lemma \ref{lemmaprincipal} except that we start with the set \[
P = \set{ p \in I^-(S) \setminus \parenthese{J^-(q_1) \cup J^-(q_2)} : d_R(I^+(p) \cap S) < 1}
\]
to create a locally f\/inite covering of $J^+(S) \setminus \parenthese{ I^-(q_1') \cup I^-(q_2') }$.
\end{proof}

The remaining of this section consists of the construction of the dif\/ferent functions.

\begin{lemma}\label{equality}
If $p \prec q$, then $\forall\, \epsilon > 0$ there exists a function $f\in \mathcal A$ such that:
\begin{itemize}
\item $\text{\rm ess} \sup g( \nabla f, \nabla f ) \leq -1$;
\item $\nabla f$ is past directed;
\item $f(q)-f(p) \geq 0$;
\item $\abs{(f(q)-f(p)) - d(p,q)} < \epsilon$.
\end{itemize}

\end{lemma}

\begin{proof}
Let us choose a smooth spacelike Cauchy surface $S$ containing $q$. Such smooth spacelike surface exists as shown by works of Bernal and S\'anchez \cite{BS03,BS04}. Then let us choose two free points~$p'$ and~$q'$ such that $p' \in I^-(p)$ and $q' \in I^+(q)$.  Because $q \in I^+(p')$ and $q\in S$ we can choose $q'$ close to $q$ such that $I^-(q') \cap J^+(S) \subset I^+(p')$.

We can apply the Lemma \ref{lemmaprincipal} to $S$, $q$ and $q'$ to get a function $f_1$ with properties:
\begin{itemize}\itemsep=0pt
\item $f_1 \in \mathcal A$;
\item $f_1 \geq 0 $;
\item $g( \nabla f_1, \nabla f_1 ) \leq -1$ and $\nabla f_1$ is past directed where $f_1 > 0$, except on a set of measure zero;
\item $f _1> 0$ on $J^+(S) \setminus I^-(q')$;
\item $f_1 = 0$ on $J^-(q)$,
\end{itemize}
and then to $S$, $p$ and $p'$ to get a function $f_2$ with properties:
\begin{itemize}\itemsep=0pt
\item $-f_2 \in \mathcal A$;
\item $f_2 \geq 0 $;
\item $g( \nabla f_2, \nabla f_2 ) \leq -1$ and $\nabla f_2$ is future directed where $f_2> 0$, except on a set of measure zero;
\item $f_2 > 0$ on $J^-(S) \setminus I^+(p')$;
\item $f_2 = 0$ on $J^+(p)$.
\end{itemize}

Then the function:
\[
f_0 = f_1 - f_2
\]
has the following properties:
\begin{itemize}\itemsep=0pt
\item $f_0\in\mathcal A$;
\item $f_0 = 0$ on the compact $J^-(q) \cap J^+(p)$;
\item the support of $\nabla f_0$ includes the set $\parenthese{J^+(S) \setminus I^-(q')} \cup  \parenthese{J^-(S) \setminus I^+(p')}$;
\item $g( \nabla f_0, \nabla f_0) \leq -1$ and $\nabla f_0$ is past directed on its support, except on a set of measure zero.
\end{itemize}
The last assertion comes from
\[
g( \nabla f_0, \nabla f_0 ) =  g( \nabla f_1, \nabla f_1 ) + g( \nabla f_2, \nabla f_2 ) -  2 g( \nabla f_1, \nabla f_2 ) \leq -1,
\]
where the f\/irst term is equal to $-1$ on the support of $f_1$ and is non-positive elsewhere, the second term is equal to~$-1$ on the support of $f_2$ and is non-positive elsewhere, and the last term is non-positive because $\nabla f_1$ and $\nabla f_2$ are not in the same orientation.

We can now def\/ine the function:
\[
f = f_0 + d_{p'}  \quad \Lequi \quad f(z) = f_0 + d(p',z)
\]
which has the following properties:
\begin{itemize}\itemsep=0pt
\item $f\in\mathcal A$;
\item $f(p)= d(p',p)$;
\item $f(q)= d(p',q)$;
\item the support of $\nabla f$ is $M$;
\item $g( \nabla f, \nabla f) \leq -1$ and $\nabla f$ is past directed on $M$, except on a set of measure zero.
\end{itemize}

To verify that the support of $\nabla f$ is $M$, we can see that $M \setminus J^+(p') \subset \parenthese{J^+(S) \setminus I^-(q')} \cup  \parenthese{J^-(S) \setminus I^+(p')} $ (recall that $I^-(q') \cap J^+(S) \subset I^+(p')$) and that the support of $d_{p'}$ is $J^+(p')$.

So now we have a function $f$ such that $f(q)-f(p) = d(p',q) - d(p',p) \geq d(p,q) \geq 0$ by the inverse triangle inequality.

Let us set the function $\alpha(p') = \parenthese{f(q) - f(p)} - d(p,q) = \parenthese{d(p',q) - d(p',p)} - d(p,q)$. $\alpha$~is a~continuous function because the distance function is continuous, and $\alpha(p) = 0$. So it is always possible to choose the initial point $p' \in I^-(p)$ such that $\abs{\alpha(p')} < \epsilon$.
\end{proof}

\begin{corollary}\label{equalityrev}
If $p \succ q$, then there exists a function $f\in \mathcal A$ such that:
\begin{itemize}\itemsep=0pt
\item $\text{\rm ess} \sup g( \nabla f, \nabla f ) \leq -1$;
\item $\nabla f$ is past directed;
\item $f(q)-f(p) \leq 0$.
\end{itemize}
In particular, $\langle f(q)-f(p) \rangle = 0$.

\end{corollary}

\begin{proof}
This is trivial by switching $p$ and $q$ in Lemma \ref{equality}.
\end{proof}

\begin{lemma}\label{zero}
If $p \nprec q$ and $q \nprec p$, then $\forall \, \epsilon > 0$ there exists a function $f\in \mathcal A$ such that:
\begin{itemize}
\item $\text{\rm ess} \sup g( \nabla f, \nabla f ) \leq -1$;
\item $\nabla f$ is past directed;
\item $\abs{f(q)-f(p)} < \epsilon$.
\end{itemize}
\end{lemma}

\begin{proof}
The case $p=q$ is trivial, so we will suppose $q\notin J^{\pm}(p)$.

Let us choose a smooth spacelike Cauchy surface $S$ containing $q$ and assume $p \in J^-(S)$ (otherwise we exchange the role of $p$ and $q$). Then we choose the following free points:
\begin{itemize}\itemsep=0pt
\item $\tilde p \in S$ such that $\tilde p \in J^+(p)$ (note: $\tilde p = p$ if $p\in S$);
\item $p' \in I^-(p)$ and $q' \in I^-(q)$ such that the sets $J^+(p') \cap S$ and $J^+(q') \cap S$ are disjoint (this is always possible because $q\in S$ and $q\notin J^{+}(p)$);
\item $\tilde p' \in I^+(\tilde p)$ and $\tilde q' \in I^+(q)$ such that $I^-(\tilde p') \cap J^+(S) \subset I^+(p')$ and $I^-(\tilde q') \cap J^+(S) \subset I^+(q')$.
\end{itemize}
Then we have an analogous situation than in Lemma \ref{equality} but with two disjoint sets
\[
\parenthese{J^+(S) \cap I^-(\tilde p')} \cup  \parenthese{J^-(S) \cap I^+(p')}\qquad\text{and}\qquad\parenthese{J^+(S) \cap I^-(\tilde q')} \cup  \parenthese{J^-(S) \cap I^+(q')}
\]
the f\/irst one containing the compact $J^-(\tilde p) \cap J^+(p)$ and the second the point $q$.

We can then apply the Corollary \ref{corollaryprincipal} a f\/irst time to $S$, $\tilde p$, $q$, $\tilde p'$ and $\tilde q'$ to get a function $f_1$ with null of past directed gradient, and a second time to $S$, $p$, $q$, $p'$ and $q'$ to get a function $f_2$ with null of future directed gradient. In the same way as in Lemma \ref{equality} we f\/ind a function:
\[
f_0 = f_1 - f_2
\]
with the following properties:
\begin{itemize}\itemsep=0pt
\item $f_0\in\mathcal A$;
\item $f_0 = 0$ on the compact $J^-(\tilde p) \cap J^+(p)$ and on $q$;
\item the support of $\nabla f_0$ includes the set $\parenthese{J^+(S) \!\setminus \! \parenthese{I^-(\tilde p') \cup I^-(\tilde q')}} \cup    \parenthese{J^-(S)\! \setminus \!\parenthese{I^+(p') \cup I^+(q') }}$;
\item $g( \nabla f_0, \nabla f_0) \leq -1$ and $\nabla f_0$ is past directed on its support, except on a set of measure zero.
\end{itemize}

Finally we def\/ine the function $f$:
\[
f = f_0 + d_{p'} + d_{q'}  \quad \Lequi \quad   f(z) = f_0 + d(p',z) + d(q',z)
\]
which has the following properties:
\begin{itemize}\itemsep=0pt
\item $f\in\mathcal A$;
\item $f(p)= d(p',p)$;
\item $f(q)= d(q',q)$;
\item the support of $\nabla f$ is $M$;
\item $g( \nabla f, \nabla f) \leq -1$ and $\nabla f$ is past directed on $M$, except on a set of measure zero.
\end{itemize}

Once more we can choose $p' \in I^-(p)$ and $q' \in I^-(q)$ such that $d(p',p) < \frac\epsilon2$, $d(q',q) < \frac\epsilon2$ and conclude that $\abs{f(q)-f(p)} < \epsilon$.
\end{proof}

Lemma \ref{zero} concludes the proof of the Theorem~\ref{mainth}.

Notes:
\begin{itemize}\itemsep=0pt
\item For every set $\mathcal A \subset C(M,\setR)$ of a.e. dif\/ferentiable functions with absolute continuity condition respected on causal paths and such that $d_p \in \mathcal A$, the construction above holds. Indeed, path integrations take place into compact sets and the f\/inal function $f$ constructed in Lemmas~\ref{equality} and~\ref{zero} is a f\/inite sum of distance functions on each compact set, so the fundamental theorem of calculus still holds. A proof of the absolute continuity of the function $d_p$ would be then a good improve of our result.
\item We think that the f\/inal function $f$ in Lemmas \ref{equality} and \ref{zero} could also be replaced by the distance $f(z) = d(S,z) - d(z,S)$ to a suitable smooth Cauchy surface $S$ (in fact, a surface containing~$p$ and such that the geodesic from~$p$ to~$q$ is maximal in Lemma \ref{equality} or con\-taining both~$p$ and~$q$ in Lemma \ref{zero}) if one can prove the existence of such surface, the a.e.\ dif\/ferentiability and the absolute continuity of the distance $d(S,z)$.
\end{itemize}

\section{Conclusion and outlook}

This paper shows the construction of a Lorentzian distance function only based on a global eikonal condition and independent of any path consideration. This is the counterpart in Lo\-rent\-zian geometry to the f\/irst step of the construction of Connes' Riemannian distance function. This is done by integrating an inverse Cauchy--Schwartz inequality along arbitrary causal paths and by constructing some suitable functions to obtain the equality.

The next step is now to f\/ind a corresponding algebraic constraint to our geometrical timelike eikonal inequality which could be extended to noncommutative structures, such as spectral triples or equivalent. A priori, this could be done in two dif\/ferent manners. A f\/irst way would be to replace the Dirac operator by a more suitable hyperbolic operator, such as the d'Alembert operator proposed by Moretti in~\cite{Mor}. The second way would be to conserve a Dirac operator but with the add of a Krein--Space structure, as shown by Strohmaier in~\cite{Stro}. In both cases, the biggest challenge would be to set a structure and a constraint that would conserve all information on Lorentzian causality.

\subsection*{Acknowledgements}

We would like to thank the anonymous referees for their useful remarks and relevant contribution to this paper.

\pdfbookmark[1]{References}{ref}
\LastPageEnding

\end{document}